# Study of the magnetic properties of the compound $M_nB_i$ using the Monte Carlo simulations

**S. Aouini, T. Sahdane, A. Mhirech, L. Bahmad[*], B. Kabouchi**

Laboratoire de Matière Condensée et Sciences Interdisciplinaires (LaMCScI),
Mohammed V, University of Rabat, Faculty of Sciences, B.P. 1014, Rabat, Morocco

**Abstract:**

The aim of this paper is to investigate the magnetic properties of the $M_nB_i$ magnetic systems by using the Monte Carlo simulation. Indeed, we analyzed the effect of the number of layers on the magnetic properties of the compound $M_nB_i$. The magnetization, the binder cumulant, the Curie temperature and the hysteresis cycle are also calculated in this work. We used the free boundary conditions to simulate the properties of the studied system, and we presented the effect of multilayer numbers on the magnetic properties of $M_nB_i$ films. Meanwhile, there are clear indications that coercive field increases drastically with the increases multilayer numbers is consistent with the experimental facts.

[*]) Corresponding authors: bahmad@fsr.ac.ma; Lahou2002@gmail.com

## 1. Introduction:

The compound $M_nB_i$ is an interesting ferromagnetic material. It has a curie temperature well above ambient temperature and appreciable coercively, which increases with increasing temperature. $M_n$ alloys generally have an anti-ferromagnetic order because they have almost filled 3D bands, but $M_nB_i$ is one of the few known ferromagnetic compounds of manganese that can be used as a permanent magnet [1-3]. In fact, $M_nB_i$ is a ferromagnetic inter-metallic structure with hexagonal structure of $N_iA_s$ type. $M_nB_i$ has been interesting due to the exceptionally high magnetic anisotropy of the low-temperature phase [4] and the favorable magneto-optical properties of the high-temperature phase [5]. It is remarkable that the coercivity of the low temperature phase increases with temperature and is much greater than that of the $N_d$-$F_e$-B magnets at higher temperatures. The low-temperature phase of $M_nB_i$ has been of great interest because of its important permanent magnetic and magnetic properties [6, 7]. The properties of $M_nB_i$ such as a high Kerr rotation [8, 9], a high transport rotational polarization and a large magnetic anisotropy perpendicular to the ambient temperature [10,11]. This compound has also led to potential applications such as magneto-optical modulation and spin-spin injectors [12]. The ferromagnetic nature of the alloys of manganese and bismuth was first reported [13, 14]. Bekier considered probable the formation of an $M_nB_i$ phase as probable between the pure manganese and the molten alloy containing 9% manganese [15-17]. Furst and Halla concluded from radiographic studies that only one compound was present with the $M_n2B_i$ structure [18, 19]. However, Montignie showed that $M_nB_i$ was the only stable compound. In other studies, the same results have been obtained by Halla and Montignie. Nevertheless, studies on fundamental and applied properties relevant to permanent magnetism have never been abandoned and this material has attracted the attention of a new generation of researchers. Hihara and Koi are studied the temperature dependence of the easy axis of magnetization in $M_nB_i$ using the nuclear magnetic resonance method [20].

In this work, we will study the compound $M_nB_i$ theoretically by using the Monte Carlo simulation to establish the magnetic properties of this material.

The outline of this paper is as follows. We describe the model and the formulations used and Monte Carlo simulation in section 2. Section 3 is dedicated to results and discussions. A brief conclusion is in given in section 4.

## 2. Model and Monte Carlo simulation:

The studied system is presented in figure 1.

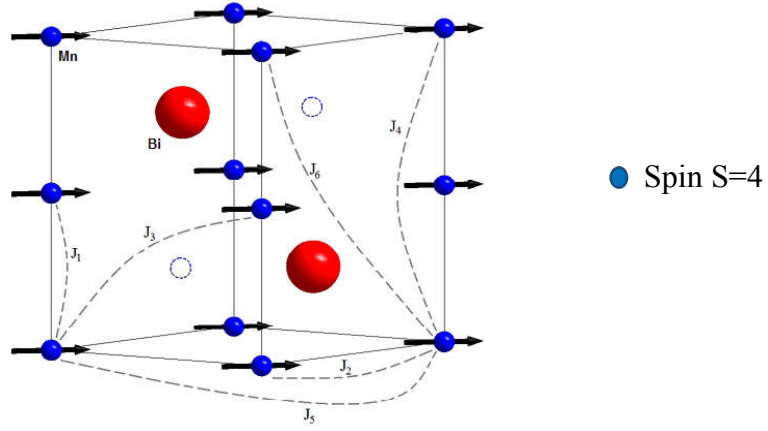

**Fig. 1:** Schematic representation of the compound $M_nB_i$. Figure reproduced from Ref. [21].

Our data were generated with $10^5$ Monte Carlo steps per spin. We take a the number of spin (N× N × nL) with N = 100, n is the number of layers and L = 10.

The Hamiltonian for this model is given by:

$$H = -J_1\sum_{\langle ij \rangle} S_i^1 S_j^1 - J_2\sum_{\langle ij \rangle} S_i^2 S_j^2 - J_3\sum_{\langle ij \rangle} S_i^3 S_j^3 - J_4\sum_{\langle ij \rangle} S_i^4 S_j^4 - J_5\sum_{\langle ij \rangle} S_i^5 S_j^5 - J_6\sum_{\langle ij \rangle} S_i^6 S_j^6 - H\sum_j S_j - \Delta\sum_j S_j^2 \quad (1)$$

With: $J_1$=4.70 me V, $J_2$= -0.61 me V, $J_3$= -1.73 me V, $J_4$= -0.12 me V, $J_5$= -1.29 me V, $J_6$= -0.63 me V [21].

The total magnetization per site:

$$M = \frac{1}{N}\sum_j S_j \quad (2)$$

The internal energy per site:

$$E = \frac{\langle H \rangle}{N} \quad (3)$$

The magnetic susceptibility is given by:

$$\chi = \beta(\langle M^2 \rangle - \langle M \rangle^2) \quad (4)$$

Where: $\beta = 1/k_B T$ with T is the absolute temperature and ($k_B = 1$) is Boltzmann's constant.

Important additional information can be may obtain by examining higher order moments of the finite size lattice probability distribution. This can be done quite effectively by considering the reduced fourth order cumulant of the order parameter (Binder 1981). Binder cumulant is defined by:

$$U_L = 1 - \frac{\langle M^4 \rangle}{3 \langle M^2 \rangle^2} \quad (5)$$

### 3. Results and discussions:

Starting from the Hamiltonian (1) of the studied system, such parameters can produce many possible configurations in different phase diagrams. By computing and comparing all possible configurations (4), we examine the phase diagrams in different planes. Indeed, the corresponding ground state phase diagrams are presented in figure 2.

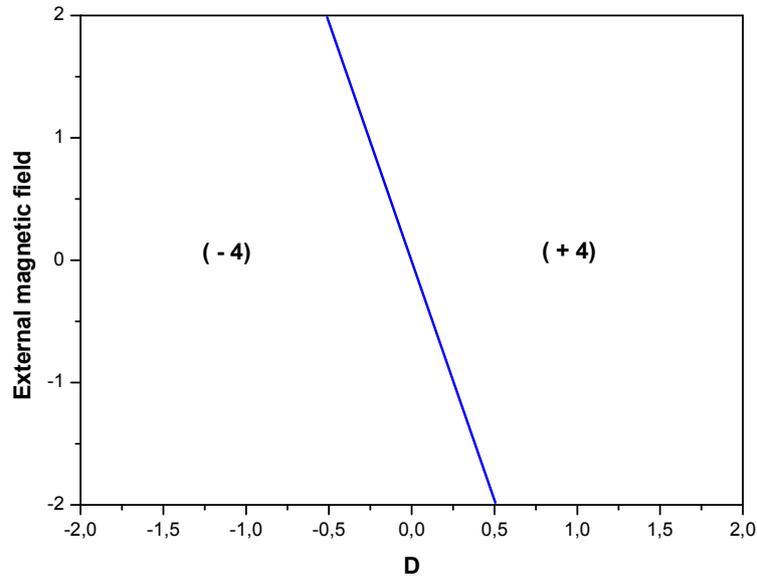

**Fig. 2:** The ground state phase diagram in the plane (D, H)

The magnetic properties of ground state of this system are studied with Monte Carlo simulations. Two configurations for the parameters of the given system correspond to the stable state. Figure 2 illustrates the phase diagram of the ground state in the plane (D, H) which the configurations (- 4) and (+4) are found stable in this figure.

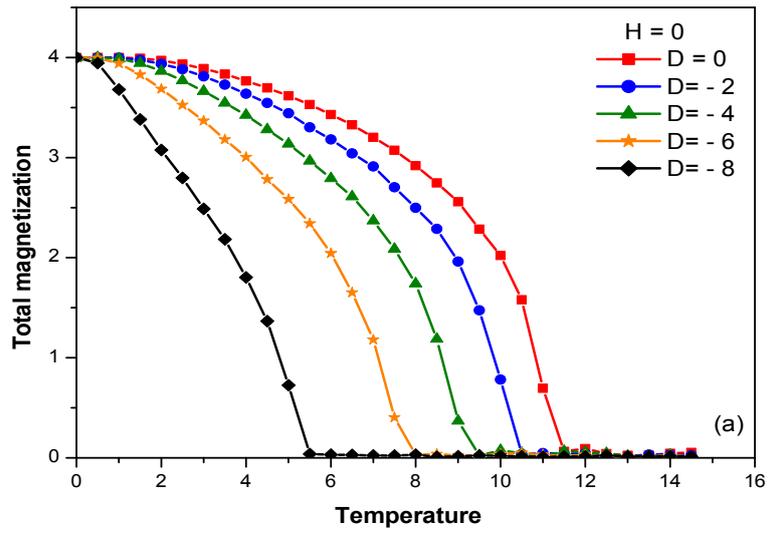

(a)

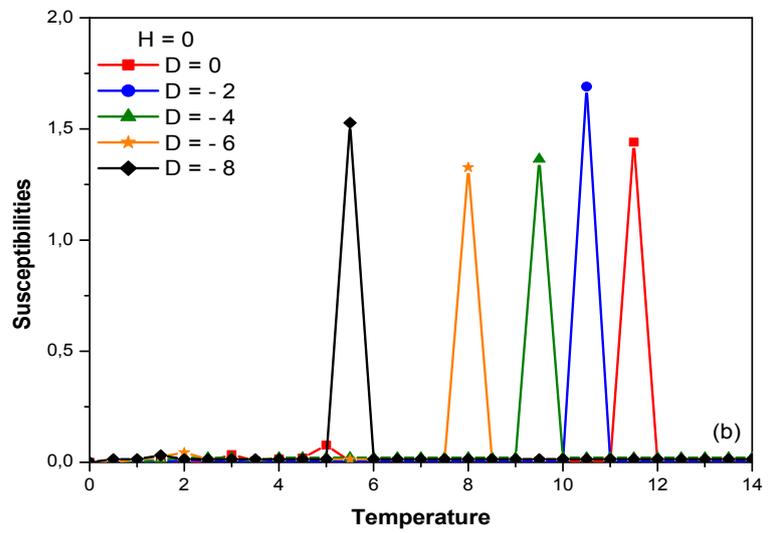

(b)

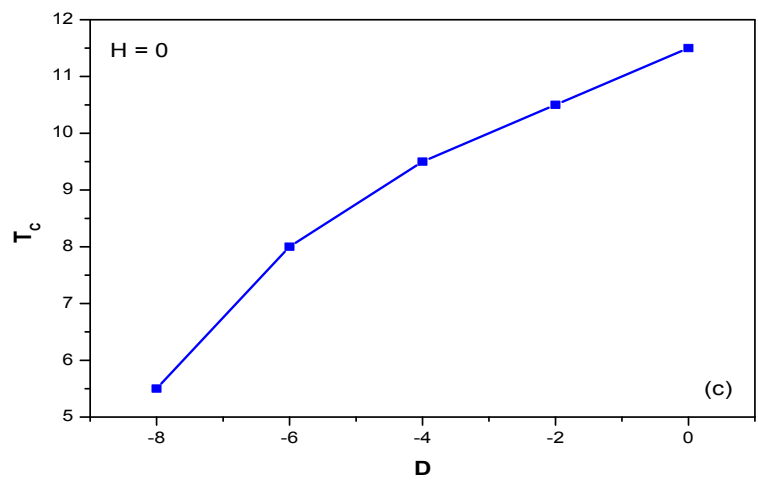

(c)

**Fig. 3: (a)** The total magnetization as a function of temperature for different values of D and H=0. **(b)** The susceptibilities as a function of temperature for different values of D and H=0. **(c)** The temperature $T_c$ as a function of the crystal field for H=0.

Figure 3(a) presents the total magnetization as a function of the temperature for different values of the crystal field in the absence of the external magnetic field (H=0). From this figure, the increasing values of the crystal field increase the critical temperature $T_c$ of this structure.

Figure 3(b) shows the susceptibilities as a function of the crystal field in the absence of the external magnetic field (H=0). We found some peaks corresponding to each value of the crystal field with different values of the temperature.

To confirm this behavior in the absence of the external magnetic field, we illustrated in figure 3(c) the increasing values of the crystal field increase the critical temperature $T_c$ of this structure.

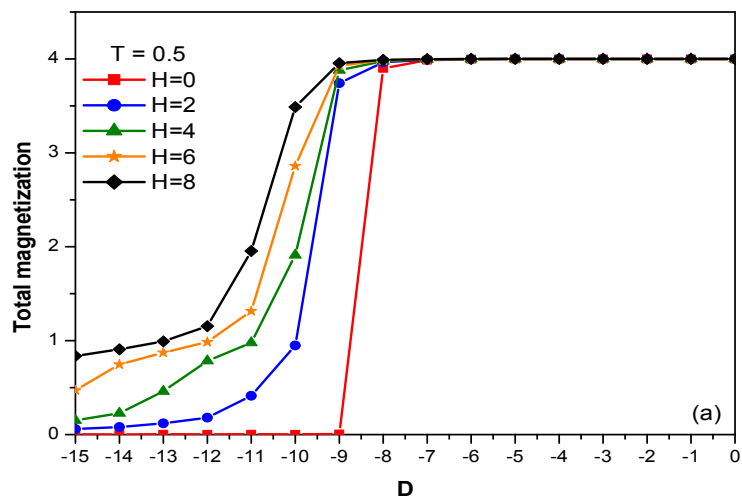

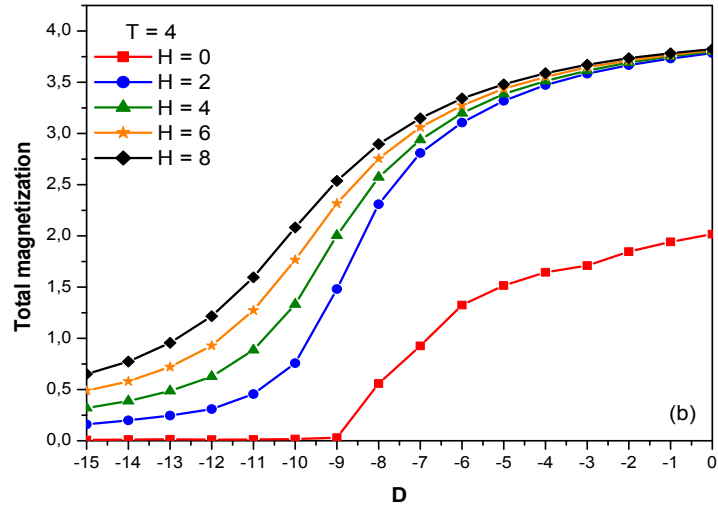

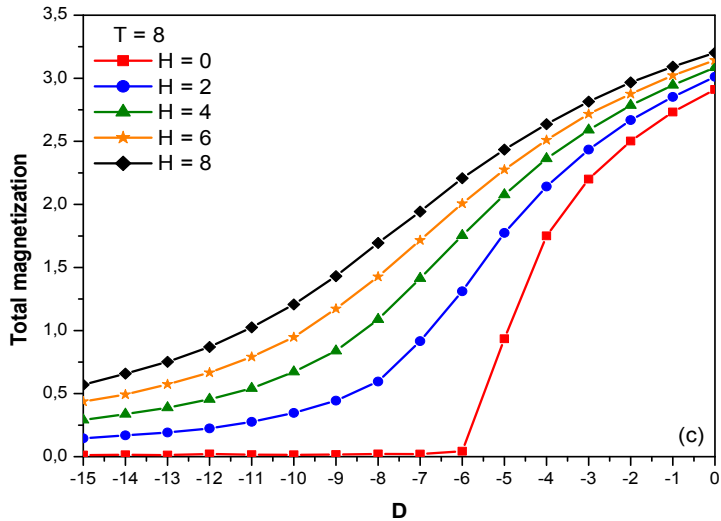

**Fig. 4:** The total magnetization as a function of the crystal field for different values of the external magnetic field (H=0, 2, 4, 6, 8) for: **(a)** T = 0.5, **(b)** T = 4 and **(c)** T = 8.

Figures 4(a)-(c) illustrate the total magnetization as a function of the crystal field with T=0.5, T=4 and T=8 respectively for different values of the external magnetic field (H=0, 2, 4, 6, 8). A more interesting behavior in this figure for crystal field, it is found that the total magnetization of this material is not affected by the external magnetic field for D >-8 and T=0.5, see figure 4(a).

For T =4 and T =8, the total magnetization increases with increasing the value of the external magnetic field for a fixed value of the crystal field, see figures 4(b).

The total magnetization decreases with decreasing the value of the crystal field in the absence of the external magnetic field (H=0), and reach to zero for D < -6, see figure 4(c).

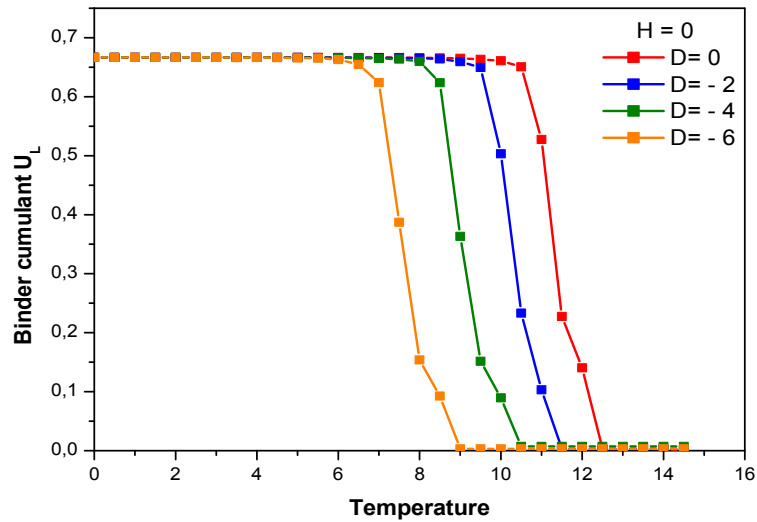

**Fig. 5:** The binder cumulant ($U_L$) as a function of the temperature for different values of crystal field (D =0, -2, -4, -6, -8) and H=0.

The curves representing the binder cumulant ($U_L$) corresponding to different values of the crystal field. This behavior is displayed in figure 5. The values obtained of the critical temperature $T_C$ agree with those in figure 3.

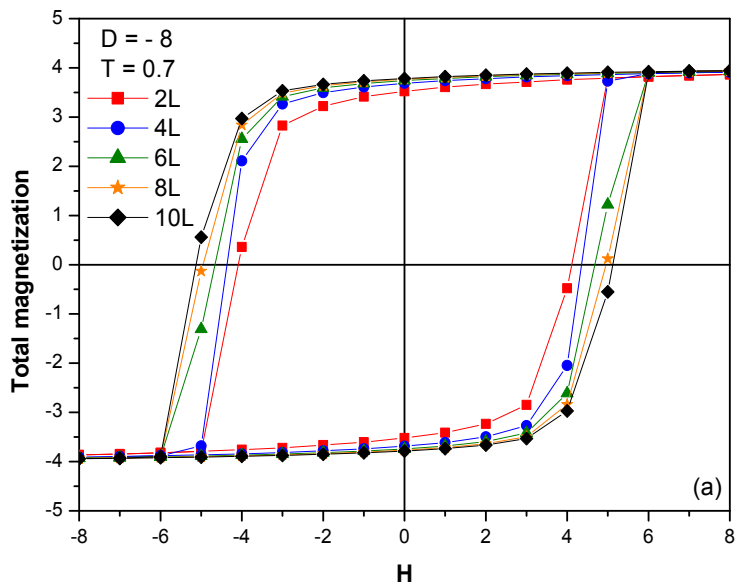

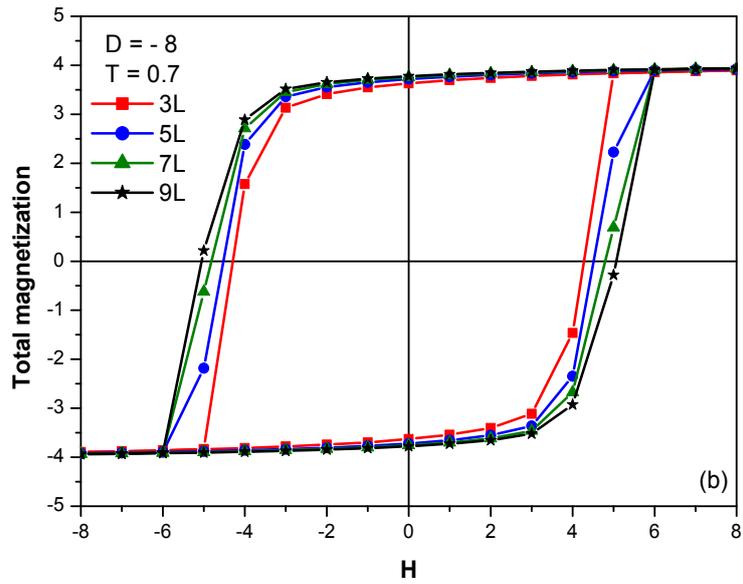

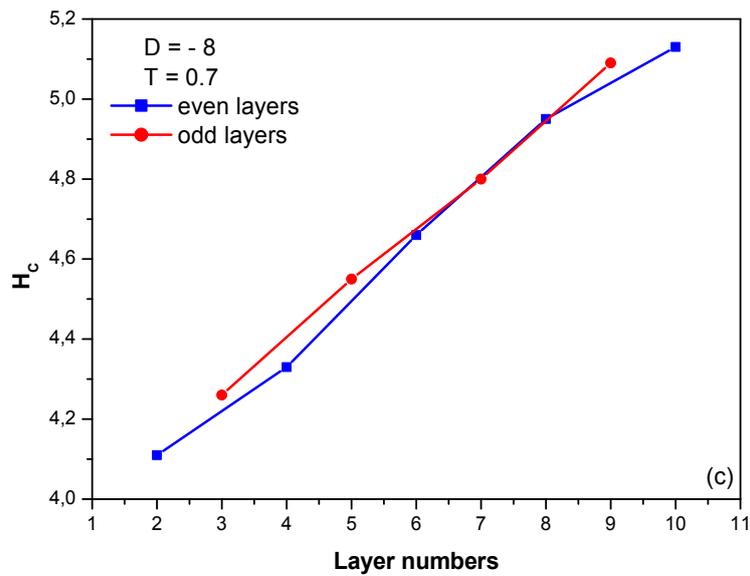

**Fig. 6:** The magnetic hysteresis loops for the multilayer $M_nB_i$ as a function of the external magnetic field H **(a)** For even layers with: T=0.7 and D=-8 **(b)** For odd layers with T=0.7 and D=-8 **(c)** The coercive magnetic field $H_c$ as a function of the layer numbers.

The magnetic hysteresis cycles for the multilayered $M_nB_i$, were obtained at the temperature T=0.7 and the crystal field D=-8. The results for even layers (2L, 4L, 6L, 8L, and 10L) and odd layers (3L, 5L, 7L, and 9L) are shown in figure 6(a) and (b), respectively. The magnetic properties of the films with even layers and odd layers show a similar dependence on the number of layers.

On the other hand, the coercive magnetic field $H_c$ is obtained from the curves, and the values are exhibited in figure 6(c). When increasing the number of layers, $H_c$ shows an increasing trend.

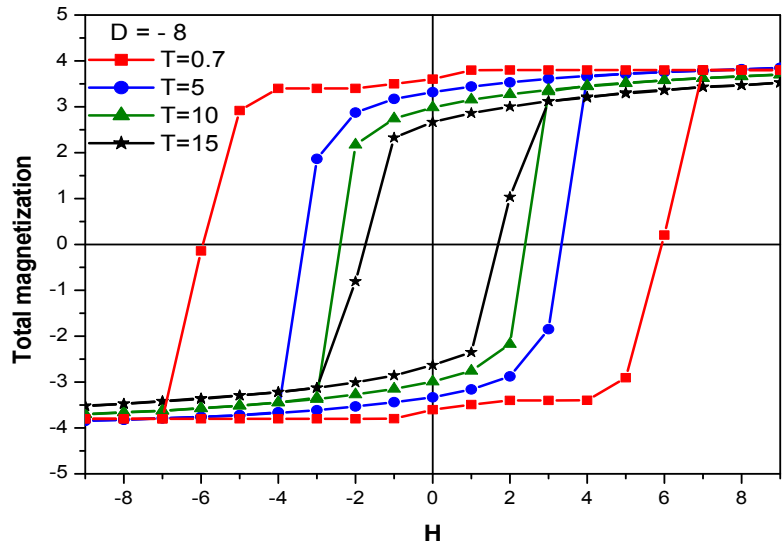

**Fig. 7:** Magnetic hysteresis cycle as a function of the external magnetic field H for different values of temperatures (T =0.7, 5, 10, 15) and D = - 8.

Our goal in figure 7 is to show the effect of temperature on the hysteresis cycles of the studied system. For fixed value of the crystal field (D = -8), we plot the behavior of the total magnetizations as a function of the external magnetic field for different values of temperature T. We observed that the region of the magnetic hysteresis cycle decreases with increasing temperature.

## 4. Conclusion

The magnetic properties of $M_nB_i$ are studied with Monte Carlo simulations. We investigated the effect of the number of layers on the magnetic properties of $M_nB_i$. We have found that the changes strongly affect the variation of magnetic properties and that the number of layers is important for obtaining superior magnetic properties. The magnetization, the binder cumulant, the Curie temperature and the hysteresis cycle are also examined in this paper. While the region of the magnetic hysteresis cycle decreases with increasing temperature. On the other hand, the coercive magnetic field $H_c$ shows an increasing trend, when increasing the number of layers. The magnetic properties of the films with even layers and odd layers show a similar dependence on the number of layers.